\documentclass[aps,prl,twocolumn,amsfonts,amssymb,amsmath,showpacs,floatfix]{revtex4}

\usepackage{graphicx,color,psfrag}
\usepackage{dcolumn}

\newcommand{\conjug}[1]{\bar{#1}}
\renewcommand{\exp}[1]{e^{#1}}
\renewcommand{\vec}[1]{\mathbf{#1}}

\newcommand{\order}{\mathcal{O}}

\newcommand{\re}{\mathop{\mathrm{Re}}}
\newcommand{\im}{\mathop{\mathrm{Im}}}

\newcommand{\sgn}{\mathop{\mathrm{sign}}}

\begin{document}
\title{Forced patterns near a {T}uring-{H}opf bifurcation}
\author{Chad M. Topaz}
\affiliation{Dept. of Mathematics and Computer Science, Macalester College, St. Paul, MN 55105, USA}
\author{Anne J. Catll{\'a}}
\affiliation{Dept. of Mathematics, Wofford College, Spartanburg, SC 29303, USA}
\date{\today}

\begin{abstract}
We study time-periodic forcing of spatially-extended patterns near a Turing-Hopf bifurcation point. A symmetry-based normal form analysis yields several predictions, including that (i) weak forcing near the intrinsic Hopf frequency enhances or suppresses the Turing amplitude by an amount that scales quadratically with the forcing strength, and (ii) the strongest effect is seen for forcing that is detuned from the Hopf frequency. To apply our results to specific models, we perform a perturbation analysis on general two-component reaction-diffusion systems, which reveals whether the forcing suppresses or enhances the spatial pattern. For the suppressing case, our results explain features of previous experiments on the CDIMA chemical reaction. However, we also find examples of the enhancing case, which has not yet been observed in experiment. Numerical simulations verify the predicted dependence on the forcing parameters.
\end{abstract}


\maketitle

Turing patterns, originally conjectured as the basis for biological morphogenesis~\cite{t1952}, arise in such diverse fields as neuroscience~\cite{b1996}, ecology~\cite{k1999}, materials science~\cite{lompke2001}, and astrophysics~\cite{ni1984}. Turing patterns were first observed experimentally in the CIMA chemical reaction~\cite{cdbd1990}; further experiments revealed a rich variety of patterns including spots, stripes, rhombs, hexagons, and superlattices~\mbox{\cite{os1991,lke1993,gos1994,bdyze2004}}. CIMA, and the closely-related CDIMA reaction, serve as convenient prototypes for other Turing systems. Additionally, these reactions are testbeds for questions of control. Spatial, temporal, and spatiotemporal forcing have been shown to induce a transition between patterns~\cite{mnmcsk2004,bm2006}, to introduce new (possibly localized) states~\cite{rmmsc2003,mpm2005}, or simply to suppress patterns~\cite{hdmze1999}.

Many systems that form Turing patterns can also display a Hopf bifurcation to time-periodic solutions. Fluid, chemical, and electrical experiments \cite{ra1985,pddddb1993,hbp1993} and theoretical studies \cite{k1980,dldb1996,mdbs1997,jbbes2001} have documented the rich dynamical behavior occurring when Turing and Hopf modes interact. If a system possesses Turing and Hopf instabilities \emph{and} is susceptible to external forcing, one may be able to manipulate and control those dynamics.

Thus, in this Letter, we study time-periodic forcing of spatially-extended systems near a Turing-Hopf bifurcation point. We give special attention to the forcing's effect on the Turing pattern. The forcing drives the Hopf mode, which in turn enhances or suppresses the Turing pattern modes to which it is coupled.  Our results are three-fold. First, we study weakly-forced Turing-Hopf bifurcations from a symmetry-based (model independent) perspective. We predict the dependence of the solutions on forcing amplitude and frequency and tie our results to experiments on the forced CDIMA system. Second, beginning with a generic two-component reaction-diffusion system, we use perturbation theory to calculate coefficients in the forced Turing-Hopf normal form. Of special interest is the coefficient of the term that couples the Hopf mode to the Turing mode, whose sign dictates whether the forcing enhances or suppresses Turing patterns. The result of the calculation can be applied to specific two-component systems such as the Lengyel-Epstein model~\cite{le1991}, the Brusselator \cite{pl1968}, and so forth to determine, as a function of the reaction kinetics, the qualitative effect of the forcing. Though previous work on the CDIMA reaction and Lengyel-Epstein model found only the suppressing effect \cite{hdmze1999,dze2001}, we find enhancement in other models. Third, we perform numerical simulations which verify the symmetry and perturbation results.

\emph{Symmetry analysis.}---Consider a spatially-extended system with a spatially homogeneous base state (zero, without loss of generality) and a codimension-two point in parameter space where steady, spatially periodic (Turing) modes and a spatially homogenous, time-periodic (Hopf) mode bifurcate simultaneously. For expository purposes, assume a simple stripe pattern. Close to the bifurcation point, let
\begin{equation}
\label{eq:modes}
\vec{u} = z_T(t_2) \vec{v}_T \exp{i\vec{q}\cdot \vec{x}} + z_H(t_2) \vec{v}_H \exp{i\omega_H t} + c.c.+\ldots,
\end{equation}
where $\vec{u}$ describes the system state (\emph{e.g.}, chemical concentrations), $z_{T,H}(t_2)$ are the slowly-varying amplitudes of the Turing and Hopf modes $\vec{v}_{T,H}$, $t_2$ is a slow time variable, $\vec{q}$ is the wave vector of the pattern with wave number $q_c=|\vec{q}|$ as determined by linear stability theory, $\omega_H$ is the Hopf frequency, and the dots represent damped modes. Assume time-periodic forcing $f(t)$. For sufficiently weak forcing, the only frequency component of $f(t)$ entering the weakly nonlinear description will be that closest to $\omega_H$. Call this frequency $\omega_f$ and its strength $f_H$. Define the detuning $\Delta = \omega_f - \omega_H$. We assume $\Delta = 0$ for now; later, we relax this assumption.

Equations for the slow-time evolution of $z_{T,H}$ must respect the spatial symmetries of the underlying system, namely translation by $\Delta \vec{x}$ ($z_T \rightarrow z_T \exp{i \vec{q} \cdot \Delta \vec{x}}$) and inversion through the origin ($z_T \rightarrow \conjug{z}_T$). Forcing breaks the temporal symmetry of the problem, but it may be recast as a parameter symmetry by incorporating the action of time translation by $\Delta t$ on the forcing function  (see, \emph{e.g.}, \cite{pts2004}) so that the full symmetry is  $(z_H,|f_H|) \rightarrow (z_H,|f_H|) \exp{i \omega_H \Delta t}$. The cubic amplitude equations are
\begin{subequations}
\label{eq:ampeqs}
\begin{eqnarray}
\dot{z}_T & = & \lambda z_T + \sgn(g_1) |z_T|^2  z_T + g_2 |z_H|^2 z_T, \\ 
\dot{z}_H & = & |f_H| + \mu z_H +  g_3 |z_H|^2  z_H + g_4 |z_T|^2 z_H.
\end{eqnarray}
\end{subequations}
Here, $\lambda,g_{1,2} \in \mathbb{R}$ and $\mu,g_{3,4} \in\mathbb{C}$.  We perform a phase shift and rescaling so that the forcing appears as $|f_H|$; we also scale such that the self-interaction coefficient of $z_T$ is $\sgn(g_1)=\pm 1$. Throughout the symmetry analysis, we assume a supercritical bifurcation to stripes, and thus set $\sgn(g_1)=-1$. For weak forcing, terms smaller than $\order(f_H)$ have no effect at leading order and are ignored.

In the absence of forcing ($f_H=0$) there are four solutions: the trivial solution, a pure Turing mode, a pure Hopf mode, and a mixed mode (see, \emph{e.g.}, \cite{dldb1996}). Weak forcing perturbs the solutions as follows.

\noindent \emph{(i) Perturbed trivial solution.} Here $z_T = 0$, $z_H \approx -|f_H|/\mu$. The physical solution~(\ref{eq:modes}) is an $\order(f_H)$ spatially homogenous oscillation with frequency $\omega_H$.

\noindent \emph{(ii) Perturbed Hopf mode.} In this case,
\begin{subequations}
\label{eq:hopfsoln}
\begin{gather}
z_T  = 0, \quad 
z_H \approx  R \exp{i \Omega t} + |f_H|(A + B \exp{2 i \Omega t}), \label{eq:hopfsol1} \\
R(\mu,g_3)  =  \sqrt{-\re(\mu) / \re(g_3)}, \\
\Omega(\mu,g_3)  =  \im(\mu) + \im(g_3) R^2.
\end{gather}
\end{subequations}
The $\order(1)$ constants $A$, $B$ are obtained by substituting (\ref{eq:hopfsoln}) into (\ref{eq:ampeqs}). The physical solution is an $\order(1)$ spatially homogeneous oscillation with frequency $\omega_H + \Omega$ superposed on $\order(f_H)$ oscillations with frequencies $\omega_H$ and $\omega_H + 2\Omega$.  The $\order(f_H)$ correction is consistent with~\mbox{\cite{eochm2000,ks2003,gpsz2008}}.

\noindent \emph{(iii) Perturbed Turing mode.} We have
\begin{subequations}
\label{eq:perturbedturing}
\begin{gather}
z_T \approx \sqrt{\lambda}\left(1+ \frac{g_2 |f_H|^2}{2 \lambda |\widetilde{\mu}|^2} \right), \\
z_H \approx -\frac{|f_H|}{\widetilde{\mu}}, \quad \widetilde{\mu}=\mu + \lambda g_4.
\end{gather}
\end{subequations}
The physical solution is an $\order(1)$ spatial pattern superposed on an $\order(f_H)$ spatially homogenous oscillation.

\noindent \emph{(iv) Perturbed mixed mode.}  In this case, $z_T$ and $z_H$ are both time-dependent:
\begin{subequations}
\label{eq:mixed}
\begin{eqnarray}
z_T & \approx & \sqrt{\lambda + g_2 \widetilde{R}^2} + |f_H| \widetilde{C}\cos( \widetilde{\Omega}t + \phi), \\
z_H & \approx & \widetilde{R} \exp{i\widetilde{\Omega}t} + |f_H| \left( \widetilde{A} + \widetilde{B} \exp{2i\widetilde{\Omega}t} \right),
\end{eqnarray}
\end{subequations}
where $\widetilde{R} = R(\widetilde{\mu},\widetilde{g}_3)$, $\widetilde{\Omega} = \Omega(\widetilde{\mu},\widetilde{g}_3)$ and $\widetilde{g}_3 = g_3 + g_2g_4$. The $\order(1)$ constants $\widetilde{A}$, $\widetilde{B}$, $\widetilde{C}$, $\phi$ are obtained by substitution. The physical solution is an $\order(1)$ spatial pattern with $\order(f_H)$ breathing at frequency $\widetilde{\Omega}$, superposed on oscillations similar to case~\emph{(ii)}.

To explore mechanisms for suppressing and enhancing Turing patterns, we consider further case \emph{(iii)}. For comparison with experimental results, we now allow detuning from $\omega_H$ ($\Delta \neq 0$), in which case (\ref{eq:perturbedturing}) becomes
\begin{equation}
\label{eq:turing_detuning}
z_T \approx \sqrt{\lambda}\left(1+ \frac{g_2 |f_H|^2}{\lambda |\widetilde{\mu}-i\Delta|^2} \right), \quad z_H \approx -\frac{|f_H|\exp{i \Delta t}}{\widetilde{\mu} - i\Delta}.
\end{equation}
The forcing's effect hinges on the sign of $g_2$. If  $g_2<0$ ($g_2>0$) the forcing suppresses (enhances) the Turing pattern.  If $g_2<0$, weak forcing reduces the pattern amplitude by a relative amount proportional to $|f_H|^2$ and shifts the domain of pattern existence from $\lambda > 0$ (unforced case) to $\lambda > -g_2 |f_H|^2 |\widetilde{\mu} - i \Delta|^{-2}  > 0$. If $g_2 > 0$, the forcing enhances the pattern and shifts the bifurcation in the opposite direction, so that a pattern appears in the forced system for $\lambda$ values where it would not exist in the unforced system. The enhancement/suppression is strongest for $\Delta_{opt} =  \im(\widetilde{\mu}) \neq 0$ and decays away from this maximum. Note that since $\widetilde{\mu}$ depends linearly on $\mu$ and $\lambda$, the shift of $\Delta_{opt}$ away from $0$ is small if we are close to the codimension-two point. Furthermore, the rate of suppression (\emph{i.e.}, the eigenvalue of the perturbed Turing mode) is approximately $-g_2 |f_H|^2 \lambda^{-1} |\widetilde{\mu}-i\Delta|^{-2}$.

The results above are consistent with experiments on the CDIMA chemical reaction \cite{hdmze1999}, which consider time-periodic forcing applied to stable Turing patterns. First, forcing was observed to suppress the pattern, corresponding to our case $g_2 < 0$. Second, suppression was strongest for $\omega_f \approx \omega_H$, in agreement with our explanation of coupling to the Hopf mode as the (indirect) control mechanism. Third, the rate of pattern suppression has a maximum near $\Delta = 0$. Finally, the numerical simulations of \cite{hdmze1999}, taken as a qualitative model of the experiment, showed that the domain of Turing pattern existence was shifted, with a critical curve in the $(|f_H|,\Delta)$ plane having a minimum near $\Delta = 0$, which follows directly from our expression for shifted domain existence.

\emph{Perturbation analysis.}---To connect the symmetry results to models, consider a general two-component reaction-diffusion system with a spatially homogeneous steady state. As in \cite{js2000}, write the governing equations as
\begin{subequations}
\begin{gather}
\label{eq:rdpde}
 \mathcal{L} \begin{pmatrix} u \\ v\end{pmatrix}= \vec{R}(u,v) + \vec{f}(t),\\
\label{eq:linearop}
\mathcal{L}\equiv\begin{pmatrix} \partial_t - a - \nabla^2 & -b \\ -c & \partial_t - d - K\nabla^2 \end{pmatrix},
\end{gather}
\end{subequations}
where $\mathcal{L}$ is the linear operator, $\vec{f}(t)$ is direct, spatially-homogeneous forcing and $\vec{R}$ contains nonlinear reaction terms. For convenience, define $\vec{R}_{2}$ ($\vec{R}_{3}$) to be the quadratic (cubic) terms in the Taylor expansion of $\vec{R}$.

From linear analysis, at the codimension-two point the parameters in (\ref{eq:linearop}) satisfy $d = -a$, $a^2 + b c < 0$, $b c < 0$, $a^2 (K+1)^2 + 4 K b c = 0$. The critical modes in (\ref{eq:modes}) are $\vec{v}_T=(-b,a-q_c^2)^T$, $\vec{v}_H=(-b, a-i\omega_H)^T$. The squared critical wave number is $|\vec{q}_c|^2 = q_c^2 = a(K-1)/(2K)$, and the squared Hopf frequency is $\omega_H^2 = -(a^2+b c) = Kq_c^4$.

To compute the coefficients in (\ref{eq:ampeqs}), we perform a multiple time scales perturbation expansion (for the case of weak forcing). The result of the perturbation expansion is (after rescaling) equation (\ref{eq:ampeqs}), including expressions for the coefficients as functions of the parameters in (\ref{eq:rdpde}). Of special interest is the sign of $g_2$, 
\begin{equation}
\sgn g_2 = \sgn\langle\vec{v}_T^{\dagger},\left(\partial\vec{R}_2\mathcal{L}^{-1} \vec{R}_2+\vec{R}_3\right)|_{\vec{u}_1}\rangle.
\end{equation}
Here $\vec{v}_T^{\dagger}=(c,-a+q_c^2)^T\exp{i\vec{q}_c\cdot\vec{x}}$, $\vec{u}_1$ is composed of the critical modes in (\ref{eq:modes}), and $\langle \vec{r},\vec{s} \rangle \equiv \conjug{\vec{r}} \cdot \vec{s}$.

\begin{figure}[t]
\centerline{
%
%
\begin{psfrags}%
\psfragscanon%
%
\psfrag{s05}[lB][lB]{\color[rgb]{0,0,0}\setlength{\tabcolsep}{0pt}\begin{tabular}{l}$A$\end{tabular}}%
\psfrag{s06}[lB][lB]{\color[rgb]{0,0,0}\setlength{\tabcolsep}{0pt}\begin{tabular}{l}$(a)$\end{tabular}}%
\psfrag{s07}[lB][lB]{\color[rgb]{0,0,0}\setlength{\tabcolsep}{0pt}\begin{tabular}{l}$g_1$\end{tabular}}%
\psfrag{s08}[lB][lB]{\color[rgb]{0,0,0}\setlength{\tabcolsep}{0pt}\begin{tabular}{l}$g_2$\end{tabular}}%
%
\psfrag{x01}[t][t]{$0$}%
\psfrag{x02}[t][t]{$0.1$}%
\psfrag{x03}[t][t]{$0.2$}%
\psfrag{x04}[t][t]{$0.3$}%
\psfrag{x05}[t][t]{$0.4$}%
\psfrag{x06}[t][t]{$0.5$}%
\psfrag{x07}[t][t]{$0.6$}%
\psfrag{x08}[t][t]{$0.7$}%
\psfrag{x09}[t][t]{$0.8$}%
\psfrag{x10}[t][t]{$0.9$}%
\psfrag{x11}[t][t]{$1$}%
\psfrag{x12}[t][t]{$8$}%
\psfrag{x13}[t][t]{$12$}%
\psfrag{x14}[t][t]{$16$}%
\psfrag{x15}[t][t]{$20$}%
%
\psfrag{v01}[r][r]{$0$}%
\psfrag{v02}[r][r]{$0.1$}%
\psfrag{v03}[r][r]{$0.2$}%
\psfrag{v04}[r][r]{$0.3$}%
\psfrag{v05}[r][r]{$0.4$}%
\psfrag{v06}[r][r]{$0.5$}%
\psfrag{v07}[r][r]{$0.6$}%
\psfrag{v08}[r][r]{$0.7$}%
\psfrag{v09}[r][r]{$0.8$}%
\psfrag{v10}[r][r]{$0.9$}%
\psfrag{v11}[r][r]{$1$}%
\psfrag{v12}[r][r]{$-6$}%
\psfrag{v13}[r][r]{$-4$}%
\psfrag{v14}[r][r]{$-2$}%
\psfrag{v15}[r][r]{$0$}%
%
\includegraphics[width=8.646cm]{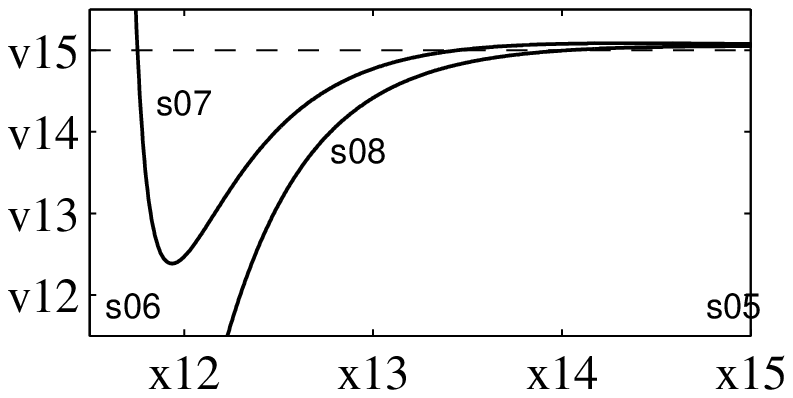}%
\end{psfrags}%
%
}
\centerline{
%
%
\begin{psfrags}%
\psfragscanon%
%
\psfrag{s05}[lB][lB]{\color[rgb]{0,0,0}\setlength{\tabcolsep}{0pt}\begin{tabular}{l}$A$\end{tabular}}%
\psfrag{s06}[lB][lB]{\color[rgb]{0,0,0}\setlength{\tabcolsep}{0pt}\begin{tabular}{l}$(b)$\end{tabular}}%
\psfrag{s07}[lB][lB]{\color[rgb]{0,0,0}\setlength{\tabcolsep}{0pt}\begin{tabular}{l}$g_1$\end{tabular}}%
\psfrag{s08}[lB][lB]{\color[rgb]{0,0,0}\setlength{\tabcolsep}{0pt}\begin{tabular}{l}$g_2$\end{tabular}}%
%
\psfrag{x01}[t][t]{$0$}%
\psfrag{x02}[t][t]{$0.1$}%
\psfrag{x03}[t][t]{$0.2$}%
\psfrag{x04}[t][t]{$0.3$}%
\psfrag{x05}[t][t]{$0.4$}%
\psfrag{x06}[t][t]{$0.5$}%
\psfrag{x07}[t][t]{$0.6$}%
\psfrag{x08}[t][t]{$0.7$}%
\psfrag{x09}[t][t]{$0.8$}%
\psfrag{x10}[t][t]{$0.9$}%
\psfrag{x11}[t][t]{$1$}%
\psfrag{x12}[t][t]{$0$}%
\psfrag{x13}[t][t]{$1$}%
\psfrag{x14}[t][t]{$2$}%
\psfrag{x15}[t][t]{$3$}%
\psfrag{x16}[t][t]{$4$}%
%
\psfrag{v01}[r][r]{$0$}%
\psfrag{v02}[r][r]{$0.1$}%
\psfrag{v03}[r][r]{$0.2$}%
\psfrag{v04}[r][r]{$0.3$}%
\psfrag{v05}[r][r]{$0.4$}%
\psfrag{v06}[r][r]{$0.5$}%
\psfrag{v07}[r][r]{$0.6$}%
\psfrag{v08}[r][r]{$0.7$}%
\psfrag{v09}[r][r]{$0.8$}%
\psfrag{v10}[r][r]{$0.9$}%
\psfrag{v11}[r][r]{$1$}%
\psfrag{v12}[r][r]{$-20$}%
\psfrag{v13}[r][r]{$0$}%
\psfrag{v14}[r][r]{$20$}%
%
\includegraphics[width=8.646cm]{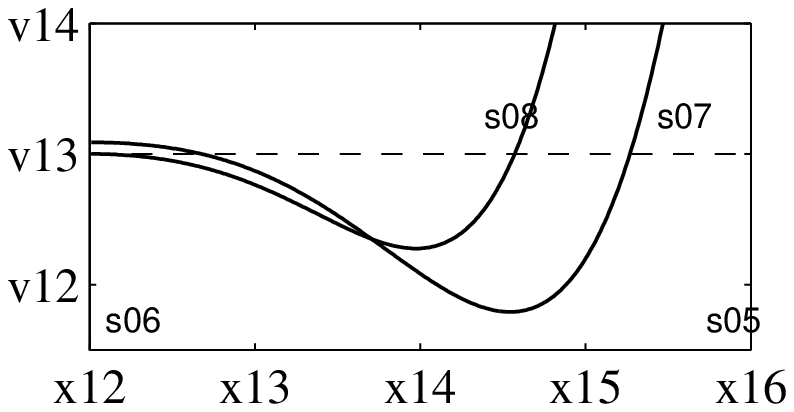}%
\end{psfrags}%
%
}
\vspace{-.15in}
\caption{Coefficients $g_{1,2}$ in (\ref{eq:ampeqs}) as computed from two reaction-diffusion models, namely the Lengyel-Epstein equations (\ref{eq:lengyelepstein}) and the Brusselator (\ref{eq:brusselator}). At the Turing-Hopf point, the coefficients in (\ref{eq:ampeqs}) depend only on the parameter $A$ in the governing equations. For each case, we focus on the interval of $A$ for which $g_1 < 0$ so that the bifurcation to a Turing pattern is supercritical. (a) In the Lengyel-Epstein equations, $g_1 < 0$ for $A \in [7.0,13.8]$. The coupling coefficient $g_2 < 0$, so that forcing suppresses the spatial pattern. (b) In the Brusselator, $g_1 < 0$ for $A \in [0.68,3.26]$. For $A \in [0.68,2.57]$, $g_2 < 0$, so that forcing suppresses the spatial pattern. For  $A \in (2.57,3.26]$, $g_2 > 0$, so that forcing actually enhances the spatial pattern.}
\label{fig:coeffs}
\end{figure}

Consider first the Lengyel-Epstein model of the CDIMA reaction with forcing  \cite{le1991,mdze1999},
\begin{subequations}
\label{eq:lengyelepstein}
\begin{eqnarray}
\dot{u} & = & A - u - 4uv(1+u^2)^{-1} - f(t) + \nabla^2 u,\\
\dot{v} & = & \sigma \left[ B \left\{u - uv(1+u^2)^{-1} + f(t) \right\} + D\nabla^2v \right]\!\!,\,
\end{eqnarray}
\end{subequations}
where $u,v$ represent the reacting chemical species and $A,B,\sigma,D$ are chemical parameters. At the codimension-two point, all coefficients in (\ref{eq:ampeqs}) can be written as functions of $A$. Fig. \ref{fig:coeffs}(a) shows $g_{1,2}(A)$. For all values of $A$ for which the Turing pattern bifurcates supercritically ($g_1<0$),  the coefficient $g_2 < 0$, and thus forcing suppresses the pattern, in agreement with only suppression having been observed in CDIMA experiments. As a second example, consider the forced Brusselator \cite{yze2004},
\begin{subequations}
\label{eq:brusselator}
\begin{eqnarray}
\dot{u} & = & A - (B+1)u + u^2v + \nabla^2 u + f(t),\\
\dot{v} & = & Bu - u^2v + D \nabla^2 u,
\end{eqnarray}
\end{subequations}
for which all coefficients in (\ref{eq:ampeqs}) can again be written as functions of $A$. As shown in Fig. \ref{fig:coeffs}(b), when $g_1 < 0$, there is one subinterval in which $g_2 > 0$, so that forcing will enhance the spatial Turing pattern.

\begin{figure}[t]
\centerline{\resizebox{\columnwidth}{!}{\includegraphics{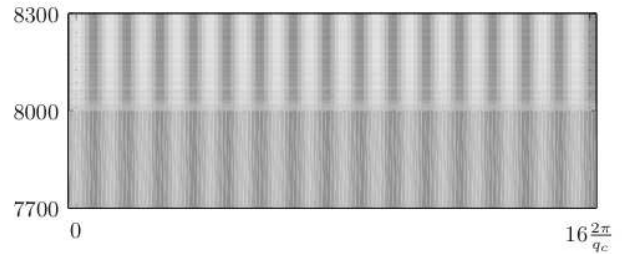}}}
\vspace{-.15in}
\caption{Simulation of the Brusselator (\ref{eq:brusselator}). Values of $u$ are indicated by shading in the $x-t$ plane. The simulation begins with $f(t) = 0$ (no forcing) and a random initial condition (not shown) which evolves to a steady state Turing pattern by $t < 8000$. At $t = 8000$, we set $f(t) = 0.01 \cos (\omega_H t)$, which enhances the Turing pattern (as seen in the sharper contrast between peaks and valleys).  The chemical parameters are $A = 3$, $B=9.998$, $D \approx 1.926$ for which $\lambda = -\re \mu = 0.001$ in (\ref{eq:ampeqs}). At the codimension-two point, $q_c \approx 1.47 $ and $\omega_H = 3$.}
\label{fig:brusselator_enhance}
\end{figure}

\begin{figure}[t]
\centerline{
%
%
\begin{psfrags}%
\psfragscanon%
%
\psfrag{s05}[lB][lB]{\color[rgb]{0,0,0}\setlength{\tabcolsep}{0pt}\begin{tabular}{l}$\log_{10}E_T$\end{tabular}}%
\psfrag{s06}[lB][lB]{\color[rgb]{0,0,0}\setlength{\tabcolsep}{0pt}\begin{tabular}{l}$\log_{10}F$\end{tabular}}%
%
\psfrag{x01}[t][t]{$0$}%
\psfrag{x02}[t][t]{$0.1$}%
\psfrag{x03}[t][t]{$0.2$}%
\psfrag{x04}[t][t]{$0.3$}%
\psfrag{x05}[t][t]{$0.4$}%
\psfrag{x06}[t][t]{$0.5$}%
\psfrag{x07}[t][t]{$0.6$}%
\psfrag{x08}[t][t]{$0.7$}%
\psfrag{x09}[t][t]{$0.8$}%
\psfrag{x10}[t][t]{$0.9$}%
\psfrag{x11}[t][t]{$1$}%
\psfrag{x12}[t][t]{$-4$}%
\psfrag{x13}[t][t]{$-3$}%
\psfrag{x14}[t][t]{$-2$}%
\psfrag{x15}[t][t]{$-1$}%
%
\psfrag{v01}[r][r]{$0$}%
\psfrag{v02}[r][r]{$0.1$}%
\psfrag{v03}[r][r]{$0.2$}%
\psfrag{v04}[r][r]{$0.3$}%
\psfrag{v05}[r][r]{$0.4$}%
\psfrag{v06}[r][r]{$0.5$}%
\psfrag{v07}[r][r]{$0.6$}%
\psfrag{v08}[r][r]{$0.7$}%
\psfrag{v09}[r][r]{$0.8$}%
\psfrag{v10}[r][r]{$0.9$}%
\psfrag{v11}[r][r]{$1$}%
\psfrag{v12}[r][r]{$-8$}%
\psfrag{v13}[r][r]{$-6$}%
\psfrag{v14}[r][r]{$-4$}%
\psfrag{v15}[r][r]{$-2$}%
\psfrag{v16}[r][r]{$0$}%
%
\includegraphics[width=8.646cm]{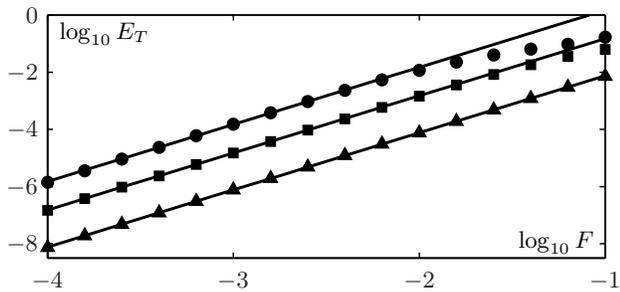}%
\end{psfrags}%
%
}
\vspace{-.15in}
\caption{Scaling of Turing pattern enhancement $E_T$ as a function of forcing strength for the Brusselator (\ref{eq:brusselator}) with forcing $f(t) = F\cos(\omega_H t)$. Symbols correspond to simulations while the lines of slope $2$ show the quadratic dependence on forcing strength predicted by (\ref{eq:turing_detuning}). For all data, $A=3$ and hence the critical wave number $q_c$ and Hopf frequency $\omega_H$ are as in Fig.~\ref{fig:brusselator_enhance}. For each data set, $\lambda = -\re \mu$ in (\ref{eq:ampeqs}). ($\bullet$) $B = 9.98$, $D \approx 1.935$ for which $\lambda = 0.01$. ($\blacksquare$) $B = 9.94$, $D \approx 1.955$ for which $\lambda = 0.03$. ($\blacktriangle$) $B = 9.8$, $D \approx 2.031$ for which $\lambda = 0.1$.}
\label{fig:forcing_scaling}
\end{figure}

\begin{figure}[t]
\vspace{-.25in}
\centerline{
%
%
\begin{psfrags}%
\psfragscanon%
%
\psfrag{s05}[lB][lB]{\color[rgb]{0,0,0}\setlength{\tabcolsep}{0pt}\begin{tabular}{l}$(a)$\end{tabular}}%
\psfrag{s06}[lB][lB]{\color[rgb]{0,0,0}\setlength{\tabcolsep}{0pt}\begin{tabular}{l}$E_T$\end{tabular}}%
\psfrag{s07}[lB][lB]{\color[rgb]{0,0,0}\setlength{\tabcolsep}{0pt}\begin{tabular}{l}$\Delta$\end{tabular}}%
\psfrag{s12}[lB][lB]{\color[rgb]{0,0,0}\setlength{\tabcolsep}{0pt}\begin{tabular}{l}$(b)$\end{tabular}}%
%
\psfrag{x01}[t][t]{$0$}%
\psfrag{x02}[t][t]{$0.1$}%
\psfrag{x03}[t][t]{$0.2$}%
\psfrag{x04}[t][t]{$0.3$}%
\psfrag{x05}[t][t]{$0.4$}%
\psfrag{x06}[t][t]{$0.5$}%
\psfrag{x07}[t][t]{$0.6$}%
\psfrag{x08}[t][t]{$0.7$}%
\psfrag{x09}[t][t]{$0.8$}%
\psfrag{x10}[t][t]{$0.9$}%
\psfrag{x11}[t][t]{$1$}%
\psfrag{x12}[t][t]{$-0.6$}%
\psfrag{x13}[t][t]{$0$}%
\psfrag{x14}[t][t]{$0.6$}%
\psfrag{x15}[t][t]{$1$}%
\psfrag{x16}[t][t]{$-1$}%
\psfrag{x17}[t][t]{$0$}%
%
\psfrag{v01}[r][r]{$0$}%
\psfrag{v02}[r][r]{$0.1$}%
\psfrag{v03}[r][r]{$0.2$}%
\psfrag{v04}[r][r]{$0.3$}%
\psfrag{v05}[r][r]{$0.4$}%
\psfrag{v06}[r][r]{$0.5$}%
\psfrag{v07}[r][r]{$0.6$}%
\psfrag{v08}[r][r]{$0.7$}%
\psfrag{v09}[r][r]{$0.8$}%
\psfrag{v10}[r][r]{$0.9$}%
\psfrag{v11}[r][r]{$1$}%
\psfrag{v12}[r][r]{$0$}%
\psfrag{v13}[r][r]{$5$}%
\psfrag{v14}[r][r]{$0$}%
\psfrag{v15}[r][r]{$2$}%
\psfrag{v16}[r][r]{$4$}%
%
\includegraphics[width=8.646cm]{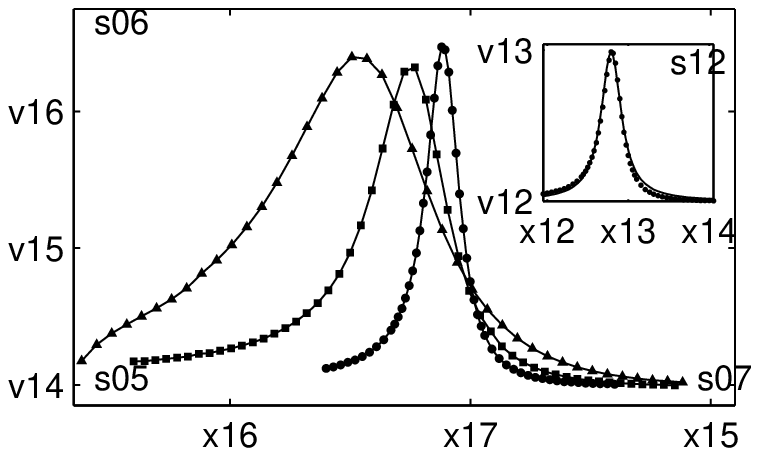}%
\end{psfrags}%
%
}
\vspace{-.15in}
\caption{Dependence of Turing pattern enhancement $E_T$ on frequency detuning $\Delta$ for the Brusselator (\ref{eq:brusselator}) with forcing $f(t) = F\cos\left((\omega_H + \Delta) t\right)$. (a) The three sets of data correspond (same symbols) to the parameters in Fig. \ref{fig:forcing_scaling}, except for the values of $F$. ($\bullet$) $F=0.001$. ($\blacksquare$) $F = 0.003$. ($\blacktriangle$) $F = 0.01$. The lines guide the eye. As predicted by (\ref{eq:turing_detuning}), there is a $\Delta_{opt}$ for which is effect is maximal and $\Delta_{opt} \to 0$ close to the codimension-two point. (b) Same data as $\bullet$ in (a), but here the line is a fit of the functional form $c_1 / (c_2 + (\Delta - c_3)^2  )$ predicted by (\ref{eq:turing_detuning}).}
\label{fig:detuning}
\end{figure}

\emph{Numerical study.}--- Fig. \ref{fig:brusselator_enhance} shows an example of Turing pattern enhancement in a simulation of (\ref{eq:brusselator}). In further simulations we quantify the Turing pattern enhancement $E_T$. To determine $E_T$, we allow the forced system to settle onto its attractor. We measure the time-averaged amplitude of the critical Turing mode, and from this subtract the corresponding value for the unforced case.  Fig.~\ref{fig:forcing_scaling} shows $E_T$ as a function of forcing strength for simple harmonic forcing and for three different sets of chemical parameters. The quadratic scaling of (\ref{eq:turing_detuning}) holds for sufficiently small forcing relative to the distance from the codimension-two point. Fig. \ref{fig:detuning}(a) shows $E_T$ vs. $\Delta$ for the same three sets of chemical parameters; as expected, there exists an optimal detuning $\Delta_{opt}$ that maximizes $E_T$, and $\Delta_{opt} \to 0$ as the system parameters approach the codimension-two point. Fig. \ref{fig:detuning}(b) verifies the functional dependence on $\Delta$ predicted by (\ref{eq:turing_detuning}).

In this Letter, we have studied the effect of time-periodic forcing on interacting Turing and Hopf instabilities. Our normal form results suggest a pattern control mechanism for spatially-extended systems with these instabilities, including chemical reaction-diffusion systems. Although the symmetry analysis is performed for weak forcing and small-amplitude patterns, it nonetheless agrees with features of the experiments in \cite{hdmze1999}.  Furthermore, we predict that the forcing may result in spatial pattern enhancement; thus far, only suppression has been observed in experiment. Nonetheless, we have demonstrated pattern enhancement with our perturbation analysis and numerical simulations. We hope that experimentalists might apply our results to look for the enhancing effect in other spatially-extended systems with Turing and Hopf instabilities.

CMT was supported by NSF grant DMS-0740484. AJC was supported by NSF VIGRE grant DMS-9983320. We thank Andrew Bernoff, Jessica Conway, Milos Dolnik, Martin Golubitsky, David Schaeffer, and Mary Silber for helpful discussions.

\bibliography{bibliography}

\end{document}